# Pain or Anxiety?

# The Health Consequences of Rising Robot Adoption in China

Qiren Liu, Sen Luo and Robert Seamans[1]

January 3, 2023

## Abstract

The rising adoption of industrial robots is radically changing the role of workers in the production process. Robots can be used for some of the more physically demanding and dangerous production work, thus reducing the possibility of worker injury. On the other hand, robots may replace workers, potentially increasing worker anxiety about their job safety. In this paper, we investigate how individual physical health and mental health outcomes vary with local exposure to robots for manufacturing workers in China. We find a link between robot exposure and better physical health of workers, particularly for younger workers and those with less education. However, we also find that robot exposure is associated with more mental stress for Chinese workers, particularly older and less educated workers.

[1] Author information: Liu: School of Economics and Statistics, Guangzhou University, Guangzhou, China; Luo: School of Economics, Jinan University, Guangzhou, China; Seamans (corresponding author): New York University, 44 West 4th Street, New York, NY 10012 USA; email: rseamans@stern.nyu.edu

## 1. Introduction

There has been a dramatic increase in robot use worldwide. As reported by Furman & Seamans (2019), annual worldwide robot sales rose from about 100,000 between 2000-2010 to over 300,000 by 2016. This increase has been particularly dramatic in China. More industrial robots have been installed in China during the past decade than in any other country. According to the *World Robotics Report 2020* released by the International Federation of Robotics (IFR), there are currently 783,000 industrial robots in operation in Chinese factories, and the rapid growth rate led Milton Guerry, president of the International Federation of Robotics, to say: "This rapid development is unique in the history of robot development."[2] This phenomenon has sparked questions about how the rapid adoption of industrial robots affects labor market outcomes, including employment, wage, and occupational structure (Acemoglu & Restrepo, 2020; Cali & Presidenti, 2021; de Vries, et al., 2020; Dixon, Hong & Wu, 2021; Giuntella & Wang, 2019; Graetz & Michaels, 2018; Maloney & Molina, 2019). Yet, other than a recent paper by Gihleb et al. (2020), little is known about the health consequences of the rapid adoption of industrial robots, especially in developing economies like China. The purpose of this paper is to fill this gap in the extant literature.

We present new evidence on the health consequences of robots for manufacturing workers in China. We focus on China for several reasons. China is now the world's largest user of industrial robots, and the use of robots in China has grown rapidly over time (Cheng, Jia, Li, & Li, 2019). Robots in China are mainly used in the manufacturing industry compared to other industries, and this gap has become extremely wide. Furthermore, China has the largest working population in the world (Fan, Lin, & Lin, 2020) and wide geographical distribution, which enables us to exploit prefecture variation to identify the impact of robot exposure on workers in manufacturing. Third, previous literature focuses almost entirely on developed countries—for example, Graetz & Michaels' (2018) cross-country study focuses exclusively on OECD countries—and there is little research on the impact of industrial robots on developing countries. As China is the largest developing country in the world, the results of our study may help managers and policymakers understand the effects of robots on workers' physical and mental health in other countries.

The health consequences of robots are an important but complex topic. On the one hand, tens of thousands of work-related injuries occur each year in the manufacturing industry,[3] which not only seriously affects the physical health of workers, but also causes economic losses to employers. Bepko & Mansalis (2016) describe the most common occupational diseases that occur in the workplace to be occupational asthma (classic asthma symptoms are cough, difficulty breathing, chest tightness, wheezing), dermatitis, and musculoskeletal disorders. Thus, one of the original purposes of robots was to replace workers in physically intensive, risky tasks which often directly result in workers' injuries and other adverse physical effects (Gihleb et al., 2020).[4] According to IFR, industrial robots are mainly used in tasks that are prone to injuries or illness, such as handling, welding, assembling, etc. Thus, robots can prevent workers from being exposed to hazardous tasks, thereby improving workers' physical health.

---

[2] See *https://ifr.org/ifr-press-releases/news/record-2.7-million-robots-work-in-factories-around-the-globe.*

On the other hand, robots may increase the mental stress on workers. Workers in industries with a high exposure to robots may worry about job security in their current workplace, and moreover may worry about increased uncertainty about labor market opportunities in general, similar to other labor market shocks (Gihleb, et al., 2020). In addition, recent popular press and academic papers have described the substitution effect of robots (Acemoglu & Restrepo, 2020; de Vries, et al., 2020; Graetz & Michaels, 2015), thereby potentially increasing workers' psychological anxiety and lowering workers' expectations for future work opportunities.[5]

We see evidence consistent with these two countervailing effects in China. The average life expectancy of Chinese citizens has increased from 71.4 years in 2010 to 77.3 years in 2019, and physical health has also gretly improved during this period. While these improvements are likely due to multiple factors, our evidence suggests that at least part of these physical health improvements may be due to increased use of robots in dangerous settings. In contrast, China has a high incidence of mental illness. Recent research shows that mental health issues, including depression, have risen over time in China (Ren et al., 2020). Our evidence suggests that anxiety about the use of robots replacing jobs may be one contributing factor.

In this paper, we explore how the use of industrial robots affects the physical and mental health of workers in China. Following Acemoglu & Restrepo (2020), we use data from the International Federation of Robots (IFR) to construct a "robot exposure" variable that measures the extent to which workers in a given industry and prefecture are exposed to robots. We combine this variable with data from two surveys, the China Health and Nutrition Survey (CHNS), from which we get individual information about the physical outcomes of workers, and the China Family Panel Studies (CFPS) from which we get mental health information about workers.

---

[3] From *https://industrytoday.com/using-robots-to-enhance-worker-safety/*.
[4] This evidence is from *https://www.safetyandhealthmagazine.com/articles/16789-robots-in-the-workplace*.
[5] A separate channel through which robots may impact health outcomes is through the use of robots in health care. Examples include the use of robots in nursing homes (Eggleston, et al., 2021) and the use of robots in surgery (e.g., Wright, 2017; Horn, et al., 2022).

We find that Chinese manufacturing workers that had a larger exposure to industrial robots experienced a significantly lower likelihood of suffering from illness or injury. Robot exposure appears linked to significant reductions of the incidence of several of the most common occupational diseases, such as rash, muscle pain, and cough. We find that the effect does not differ by gender, but the effect is most pronounced for younger workers and those with less education. However, we also find that greater robot exposure is associated with worse mental health, particularly for lower educated and older workers.

Our paper complelements recent papers by Abeliansky & Beulmann (2019) and Gihleb et al. (2020).[6] Abeliansky & Beulmann (2019) study German manufacturing workers and find an increase in robot intensity is associated with a decrease in mental health. Gihleb et al. (2020) studies the effect of robot exposure on US and German manufacturing workers and find that increased robot exposure is associated with decreased workplace-related injuries and an increase in alcohol and drug-related deaths and mental health problems. Our paper complements Gihleb et al. by utilizing several Chinese datasets to study the effects of robot exposure on health outcomes. At a high level, we find similar effects (robot exposure is associated with fewer physical health problems but more mental health problems), suggesting that robots affect workers in developing countries much in the same way as they do in developed countries. As described above, we also investigage several types of heterogeneity including worker age, education and gender that further complements the paper by Gihleb et al. (2020).

The rest of the paper proceeds as follows. Section 2 describes our main data sources and dependent and independent variables. Section 3 describes our empirical strategy and estimation sample. Section 4 provides the main empirical results and explores various sources of heterogeneity. Section 5 concludes.

## 2. Data and measures[7]

---

[6] Abeliansky & Beulmann (2019) study "robot intensity" which is calculated using a perpetual inventory method that tracks the shipments of robots into different industries in Germany. Gihleb et al. (2020) study "robot exposure" which is calculated similarly as in Acemoglu & Restrepo (2020), and as we do in this paper.
[7] Data and replication code to be provided on publication.

In this section, we describe our main sources of data for robots, workers' physical health and mental health.

*2.1. Robot data*

We use data on robots from the International Federation of Robots (IFR), which provides of counts of new robots installed by industry, country, and year.[8] IFR data has been used in many other papers on the economic effects of robots including Graetz & Michaels (2018) and Acemoglu & Restrepo (2020). Within manufacturing, we have data on the use of robots for 10 industries: food and beverages, textiles, wood and furniture, paper, plastic and chemical products, glass ceramics stone mineral products, metal, electrical electronics, automotive, other vehicles. We use this classification throughout and refer to it as the "IFR industries."

We use prefecture and industry level employment data from the China Annual Survey of Industrial Firms (ASIF), which is conducted by the Chinese government's National Bureau of Statistics, and has been widely used in studies for China (e.g., Hsieh & Klenow, 2009; Song, Storesletten, & Zilibotti, 2011). We match China's industries with IFR industries by referring to China's industrial classification for national economic activities (NEA), and aggregate the numbers of employees at the firm level provided by ASIF to the prefecture-level for each industry. We then use this data together with the industry-year level data on robots from IFR to construct our main independent variable—*Robot Exposure*—following the approach in Acemoglu & Restrepo (2020) and Giuntella & Wang (2019). Specifically, we first calculate the ratio of robots to employed workers in IFR industry *s* at the national level and then multiply it by the prefecture's baseline employment share in industry *s*, and then sum over all industries in the prefecture. Formally

$$Robot\_exposure_{c,t}^{CHN} = \sum_{s \in S} l_{c,s}^{2000} \left( \frac{Robots_{s,t}^{CHN}}{Employment_{s,2000}^{CHN}} \right) \qquad (1)$$

where $l_{c,s}^{2000}$ is the 2000 share of prefecture *c*'s employment in IFR industry *s*; $Robots_{s,t}^{CHN}$ is the stock of industrial robots in use in China (CHN) in IFR industry *s* and year *t*; and $Employment_{s,2000}^{CHN}$ is the total number of workers (in thousands) employed in industry *s* in 2000. As mentioned above, the number of works in industry *s* and prefecture *c* in 2000 are computed using aggregate data at the firm-level from ASIF. We use the superscript CHN to indicate that we calculate these figures for China.

---

[8] Considering that most industrial robots can be used for 12-15 years (IFR, 2014), we follow Fan, Hu, and Tang (2020) and compute the operational stock of for each industry.

As explained in more detail in a later section, we also construct a similar measure for Japan (JPN) that we use as an instrument.

*2.2. China Health and Nutrition Survey*

Data on workers' physical health comes from China Health and Nutrition Survey (CHNS) and Fan, Lin, et al. (2020). According to the description of the data on the official website of CHNS, it is an individual-level longitudinal survey conducted by the Carolina Population Center at the University of North Carolina at Chapel Hill and the National Institute for Nutrition and Health (NINH, former National Institute of Nutrition and Food Safety) at the Chinese Center for Disease Control and Prevention (CCDC). The survey intends to study how the social and economic transformation of Chinese society is affecting the health and nutritional status of its population. The dataset provides detailed indicators of illness, injury status, health history, family life, etc, and has been widely used for research in the fields of health economics and demographic economics (e.g., Chen, 2006; Wang, 2011). We use data for all available years starting in 2000 (2000, 2004, 2006, 2009, 2011, and 2015). We use the information provided by Fan, Lin, et al. (2020) to match the CHNS data to the prefecture level. Although their dataset covers only 52 prefectures from 11 Chinese provinces, the multistage, random cluster process method used in the CHNS sample means our estimates likely reflect the national average.

Our measures of physical health are binary variables provided by the CHNS. These are *Illness or injury*, *Muscle pain*, *Rash*, *Cough*. If the respondent answered "yes" to having one of these ailments in the past 4 weeks, the indicator takes the value of "1", and if not, the value is "0". Table 1 reports summary statistics on the main variables used in our analysis of physical health outcomes.

*2.3. China Family Panel Studies Survey*

Data on workers' mental health comes from China Family Panel Studies (CFPS) survey for the years 2010, 2012, 2014 and 2016. As described by Xu & Wang (2022), CFPS measures individual mental health using scales of the Center for the Epidemiological Studies of Depression (CES-D). Interviewees are asked to answer "the frequency of occurrence of each item during the past week" to reflect their mental health according to the CFPS questionnaire. However, the number of items and the description of each item is not consistent in different survey waves. To address the inconsistency, we adopt a similar procedure as Xu & Wang (2022) by carrying out the following steps: First, we reverse the scales of positive items to ensure that a higher scale indicates worse psychological anxiety throughout the items. Secondly, the description of the items in 2010 and 2014 differ from that in 2012 and 2016, and the number of items is also different. Therefore, we calculate the average value to harmonize the variation in item numbers. After the above steps, the mental health index is no longer a categorical variable, but a continuous real value that ranges between 1 to 5, with higher values indicating worse individual psychological status. We call this variable *Mental Stress*. Table 2 reports summary statistics on *Mental Stress* and other variables used in our analysis of mental health outcomes.

It is useful to note a few differences across the samples used for physical health and mental health outcomes. First, the data on physical health outcomes covers a much longer time period (1989 to 2015), whereas the data on mental health is more recent (2010 to 2016). As a result, average *Robot Exposure* is much larger in the mental health sample, reflecting the recent rapid adoption of robots in China. Second, none of the years overlap between the two surveys. The physical health survey is mostly conducted on odd years whereas the mental health survey is conducted on even years. Third, while we include some similar demographic variables across the two samples, we also include variables that are not common to both, such as health insurance, disease history, smoking and other variables that have been shown to be correlated with physical health outcomes (Fan et al., 2020) and hours worked (Berniell et al., 2020) and house prices (Xu & Wang, 2022) that have been shown to be correlated with mental health outcomes.

## 3. Empirical strategy and estimation sample

### 3.1. Empirical strategy

We aim to study the health consequences of rising robot adoption in China. To this end, we link workers' physical health status from the CHNS and mental health status from the CFPS with our measure of robot exposure at the prefecture-level to investigate how these health outcomes have changed for manufacturing workers. We employ the following empirical specification:

$$Y_{ict} = \alpha + \beta Robot\_exposure_{ct}^{CHN} + \delta X_{it} + \gamma X_{ht} + \rho X_{ct} + \theta_t + \vartheta_c + \varepsilon_{ict} \quad (2)$$

where the subscript $ict$ denotes the manufacturing worker $i$ located in prefecture $c$ in a given year $t$. $Y_{ict}$ is a physical or mental health outcome for individual $i$ in prefecture $c$ in year $t$. $Robot\_exposure_{c,t}^{CHN}$ denotes the robot exposure of prefecture $c$ in year $t$ as described in Eq. (1). The coefficient of interest "$\beta$" measures the average effects of rising robot adoption on the probability of reporting a physical or mental health issue. $X_{it}$ is a comprehensive set of individual characteristics that may affect one's health conditions, including gender, years of education, age, age square and disease history. We also include fixed effects for worker occupation type and employer ownership type to account for differences in the type of workers' occupations and the type of companies employed, which others have shown can have a large impact on workers' physical health (Fan et al., 2020). $X_{ht}$ denotes a set of household characteristics, including household size, household income, drinking water conditions, bathroom conditions, etc. Similarly, we control for a set of health-related factors at the prefecture-level, denoted by $X_{ct}$, such as population, the share of employment in the manufacturing sector, etc. Since it has been documented that input tariff shocks (Fan, et al., 2020), air quality (Xie et al., 2016), water pollution (Wang & Yang, 2016) can impact individual health, we also control for import tariffs, manufacturing tariffs, air quality index, and water pollution at the prefecture level. We obtain the prefecture level data from the *China City Statistical Yearbook*.

$\theta_t$ are year fixed effects, which captures common macro shocks to all individuals across years. We also include prefecture fixed effects, $\vartheta_c$, to control for all time-invariant differences across prefectures in our regressions. $\varepsilon_{ict}$ is the error term. We cluster standard errors at the prefecture level in our main specification to allow for potential correlations between individuals within the prefecture.

*3.2. Estimation sample*

In our regressions, we restrict the sample to the group of manufacturing workers who are in the legal working age group, i.e., the group of male workers aged 16-60 and female workers aged 16-55, according to the Labor Law of the People's Republic of China. In Appendix Table A1, we can see the 13 individual occupation types provided by the CHNS questionnaire, we only choose occupations "6" and "7" which are considered to be related to manufacturing workers, following Fan, et al. (2020).

**4. Empirical results**

*4.1. Physical health*

We start by investigating the link between robot exposure and physical health. We estimate Eq. (2) using a linear probability regression model (results are robust to use of a probit model, as shown in Appendix Table A2). The dependent variables are *Illness or Injury*, our main measure of worker health condition, as well as *Muscle Pain*, *Rash*, and *Cough*. The main independent variable of interest is *Robot Exposure*. We present the results in Table 3. Column (1) shows the results of regressing *Illness or Injury* on *Robot Exposure* and the full set of control variables. The coefficient is negative and significant, indicating a negative relationship between robot exposure and the probability of reporting an illness or injury. More specifically, the coefficient implies that a one-standard-deviation increase in our prefecture-level measure of robot exposure is associated with an approximately 0.75% reduction in the probability of experiencing illness or injury.[9] In Columns (2), (3) and (4) we investigate the link between robot exposure and *Muscle Pain*, *Rash*, and *Cough*, respectively. The coefficient on robot exposure is negative and statistically significant in all of these cases as well.

Next, we investigate heterogeneous effects on the link between robot exposure and illness or injury. To do this, we re-estimate equation (2) on several subsamples of the data. We present these results in Table 4. In columns (1) and (2) we split the sample by female and male respondents, respectively. The coefficient on robot exposure is negative in both cases, though only statistically significant in the case of men. The magnitude of the coefficient is similar across each sub-sample, leading us to conclude that there is little difference in the effect by gender.

Next we examine whether the effects vary across education levels. For this purpose, we separate the analysis into two subgroups: "low education" is defined as individuals with less than or equal to 12 years of education (i.e., high school graduation or less) and "high education" is defined as those with more than 12 years of education. This is based on the hypothesis that industrial robots mainly substitute for physically demanding and hazardous tasks that are typically done by workers with less education, nudging these workers towards occupations that are lower risks.[10] The results in columns (3) and (4) of Table 3 show that robot exposure is associated with a statistically significant reduction in the likelihood of illness or injury among low educated workers, while the effect on highly educated workers is positive but not significant.

Finally, we examine whether the use of industrial robots has different effects on workers of different ages. The mean age of the sample is used as the cut-off (see Table 1 for descriptive statistics), and the group of workers under the age of 42 is defined as "young workers" and the group of workers aged 42 and above is defined as "old workers", and the regression results are presented in columns (5) and (6) of Table 3. The results show that robot exposure is associated with a statistically significant reduction in the likelihood of illness or injury among young workers, while the effect on old workers is negative but not significant. A possible explanation is that younger workers more typically perform work that is physically demanding or hazardous, whereas older workers are less likely to perform such work.

---

[9] This probability value is obtained by multiplying the sample standard deviation of the robot_exposure (0.00116) by the estimated coefficient (-6.393).

[10] See for example: [6 Ways Robots Can Be Used to Address Unsafe Working Conditions (ecorobotics.com)](#) and [Robots Keep Workers off Dangerous Tasks | designnews.com](#)

*4.2. Mental health*

To investigate the link between robot exposure and mental health we estimate Eq. (2) using a linear probability regression model where the dependent variable is *Mental Stress*. The main independent variable of interest is *Robot Exposure*. Column (1) shows the results of regressing *Mental Stress* on *Robot Exposure* and the full set of control variables. The coefficient on *Robot Exposure* is positive and significant, suggesting that as robot exposure increases, so too does mental stress.

We next investigate several heterogeneous effects. We first investigate differences between female and male workers in columns (2) and (3). The coefficient on *Robot Exposure* is positive for female and positive and significant for male. The coefficient for the female subsample is sufficiently noisy that we cannot rule out that it is the same as that for male.

Next, we consider the "low education" and "high education" subsamples that we used in the analysis of physical health in Table 4 above. Column (4) provides the results for low education subsample. The coefficient on *Robot Exposure* is positive and significant. Column (5) provides the results for the high education subsample. The coefficient on *Robot Exposure* is positive but not statistically significant. However, the coefficient is sufficiently noisy that we cannot rule out that it is the same across the two subsamples.

Finally, we examine whether there are different effects on workers of different ages, as we did in Table 4. The results show that robot exposure is associated with a positive but not statistically significant increase in mental stress for young workers, and a positive and statistically significant increase in mental stress for old workers, with a coefficient that is about twice as large as that for young workers. A possible explanation is that older workers have less time to learn new skills and switch to a new occupation if their existing occupation is automated by a robot, and hence older workers experience more stress than younger workers when faced with more robot exposure.

*4.3. Robustness tests*

Our research design links changes in local robot exposure to changes in worker physical and mental health outcomes. The decision of firms in an industry to adopt robots or not is endogenous. We have therefore been careful not to use causal language when describing our findings. As a next step, we use an instrumental variables approach to study the link between robot exposure and health outcomes. To do this, we follow Acemoglu & Restrepo (2020) by instrumenting adoption of robots in Chinese industries in a given year with the adoption of robots in the same industries but in a different country. Acemoglu & Restrepo (2020) study the link between US manufacturing industry exposure to robots, and choose as their instrument European manufacturing industry robot adoption. In our case, we use robot adoption in Japanese manufacturing as an instrument. More specifically, we estimate the following equation:

$$Robot\_exposure_{c,t}^{IV} = \sum_{s \in S} l_{c,s}^{2000} \left( \frac{Robots_{s,t}^{Japan}}{Employment_{s,2000}^{CHN}} \right) \quad (3)$$

where $Robots_{s,t}^{Japan}$ denotes Japanese industrial robots used in industry *s* in the year *t*. The remaining indicators are interpreted as described in Eq. (1). In this paper, a two-stage least squares regression of Eq. (2) on instrumental variables is performed, and the results are reported in Appendix Table A3. The results are consistent with those described in the sections above.

**5. Conclusions**

Over the past two decades, Chinese adoption of industrial robots has increased dramatically. Recent studies find that robot adoption likely has an impact on the labor market, including employment, wage, productivity, etc. Our study complements this literature by examining the physical and mental health consequences of robot adoption in China.

Using individual-level data from CHNS and Fan et al. (2020) between 2000 and 2015, coupled with measures of robot exposure (following Acemoglu & Restrepo, 2020; Giuntella & Wang, 2019), we find that robot exposure has a positive impact on workers' physical health. Using data from CFPS between 2010 and 2016, we find evidence that workers respond to such an increase in robot exposure by higher mental stress. These outcomes are particularly pronounced for younger workers and lower educated workers.

    Overall, our findings highlight the complex relationship between the adoption of industrial robots and the health effects of workers in the sectors that are most exposed to robots. Compared with existing studies, our paper emphasizes the health consequence of this technology shock in developing countries and points out the large negative impact of this technological change on the mental health of workers, despite improving their physical well being.

Table 1: Summary Statistics On Dataset of Physical Health Outcomes

| Variable | Obs | Mean | Std.Dev. | Min | Max |
|---|---|---|---|---|---|
| Robot Exposure | 3,578 | 0.000358 | 0.00116 | 0 | 0.0106 |
| Illness or Injury | 3,567 | 0.0956 | 0.294 | 0 | 1 |
| Muscle Pain | 3,578 | 0.0235 | 0.151 | 0 | 1 |
| Rash | 2,896 | 0.00483 | 0.0694 | 0 | 1 |
| Cough | 2,899 | 0.0676 | 0.251 | 0 | 1 |
| Gender | 3,578 | 0.345 | 0.475 | 0 | 1 |
| Age | 3,578 | 38.71 | 10.05 | 16 | 60 |
| Age^2 | 3,578 | 1599 | 777.0 | 256 | 3600 |
| Years Education | 3,473 | 9.291 | 3.340 | 0 | 21 |
| Household Size | 3,575 | 3.970 | 1.397 | 1 | 11 |
| Household Income | 3,575 | 9666 | 13043 | -11867 | 248500 |
| Manufacturing Industry Output | 3,578 | 47.62 | 9.749 | 20.21 | 70.74 |
| Population | 3,578 | 546.7 | 202.8 | 142.4 | 1010 |
| Residential Status | 3,578 | 0.575 | 0.494 | 0 | 1 |
| Health Insurance | 3,578 | 0.0763 | 0.266 | 0 | 1 |
| Disease History | 3,561 | 0.0517 | 0.221 | 0 | 1 |
| Smoking | 3,578 | 0.00838 | 0.0912 | 0 | 1 |
| Log (Air Quality) | 3,578 | 4.542 | 0.401 | 3.151 | 5.260 |
| Water Pollution | 3,578 | 1.400 | 0.691 | 0.113 | 2.658 |
| Input Tariffs | 3,578 | 0.0981 | 0.0289 | 0.0511 | 0.180 |

Sources: China National Health Survey, China City Statistical Yearbook, International Federation of Robotics

Table 2: Summary Statistics On Dataset of Mental Health Outcomes

| Variable | Obs | Mean | Std.Dev. | Min | Max |
|---|---|---|---|---|---|
| Robot Exposure | 5,334 | 0.00318 | 0.00416 | 0.000223 | 0.0594 |
| Mental Stress | 4,985 | 1.730 | 0.532 | 1 | 5 |
| Gender | 5,334 | 0.583 | 0.493 | 0 | 1 |
| Age | 5,334 | 37.28 | 10.51 | 16 | 60 |
| Age^2 | 5,334 | 1500 | 795.8 | 256 | 3600 |
| Years Education | 5,215 | 3.074 | 1.211 | 1 | 7 |
| Household Size | 5,334 | 4.254 | 1.858 | 1 | 17 |
| Household Income | 5,142 | 72302 | 185428 | 10 | 1.140e+07 |
| Manufacturing Industry Output | 5,334 | 48.98 | 8.695 | 19.26 | 81.82 |
| Population | 5,334 | 553.9 | 302.5 | 19.80 | 1450 |
| Urban | 5,306 | 0.559 | 0.497 | 0 | 1 |
| Married | 5,334 | 1.922 | 0.576 | 1 | 5 |
| Hours Worked per Day | 2,875 | 8.413 | 2.367 | 0.0143 | 24 |
| Libraries | 5,328 | 442.5 | 862.7 | 11.50 | 7676 |
| House Price | 5,333 | 100000 | 238269 | 0 | 4.601e+06 |

Sources: China CFPS Survey, China City Statistical Yearbook, International Federation of Robotics

Table 3: Effect of Robot Exposure on Physical Health (OLS)

| | (1) | (2) | (3) | (4) |
|---|---|---|---|---|
| *Dependent variable:* | *Illness or Injury* | *Muscle Pain* | *Rash* | *Cough* |
| Robot Exposure | -6.393*** | -3.221*** | -1.354*** | -8.704*** |
| | (2.109) | (0.862) | (0.446) | (2.161) |
| Observations | 3,361 | 3,370 | 2,805 | 2,808 |
| R-squared | 0.099 | 0.038 | 0.030 | 0.070 |
| Demographic Controls | YES | YES | YES | YES |
| Year FE | YES | YES | YES | YES |
| Prefecture FE | YES | YES | YES | YES |
| Primary occuption FE | YES | YES | YES | YES |
| Employer ownership FE | YES | YES | YES | YES |

Robust standard errors in parentheses

*** p<0.01, ** p<0.05, * p<0.1

Table 4: Heterogeneous Effects of Robot Exposure on Physical Health

|  | (1) | (2) | (3) | (4) | (5) | (6) |
|---|---|---|---|---|---|---|
|  | *Dependent variable: Illness or Injury* | | | | | |
| Subsample | Female | Male | Low Educ | High Educ | Young | Old |
|  |  |  |  |  |  |  |
| Robot Exposure | -8.101 | -6.307** | -7.223*** | 8.061 | -6.696*** | -1.713 |
|  | (5.351) | (2.639) | (2.368) | (7.753) | (2.271) | (3.461) |
|  |  |  |  |  |  |  |
| Observations | 1,157 | 2,204 | 3,036 | 325 | 2,030 | 1,331 |
| R-squared | 0.154 | 0.098 | 0.103 | 0.263 | 0.121 | 0.116 |
| Demographic Controls | YES | YES | YES | YES | YES | YES |
| Year FE | YES | YES | YES | YES | YES | YES |
| Prefecture FE | YES | YES | YES | YES | YES | YES |
| Primary occuption FE | YES | YES | YES | YES | YES | YES |
| Employer ownership FE | YES | YES | YES | YES | YES | YES |

Robust standard errors in parentheses

*** p<0.01, ** p<0.05, * p<0.1

Table 5: Effects of Robot Exposure on Mental Health (OLS)

|  | (1) | (2) | (3) | (4) | (5) | (6) | (7) |
|---|---|---|---|---|---|---|---|
| *Dependent variable: Mental Stress* | | | | | | | |
| Subsample: | Full Sample | Female | Male | Low Educ | High Educ | Young | Old |
| Robot Exposure | 10.904* | 16.414 | 7.880** | 10.826* | 5.880 | 7.914 | 19.305*** |
|  | (6.136) | (12.320) | (3.103) | (6.373) | (35.966) | (5.903) | (4.773) |
| Observations | 2,371 | 950 | 1,407 | 2,064 | 274 | 1,514 | 834 |
| R-squared | 0.278 | 0.317 | 0.281 | 0.284 | 0.382 | 0.301 | 0.331 |
| Demographic Controls | YES | YES | YES | YES | YES | YES | YES |
| Year FE | YES | YES | YES | YES | YES | YES | YES |
| Prefecture FE | YES | YES | YES | YES | YES | YES | YES |

Robust standard errors in parentheses

*** $p<0.01$, ** $p<0.05$, * $p<0.1$

# Appendix Tables

Table A1
Occupations provided by CHNS

| Occupation numbers | Definitions |
| --- | --- |
| 01 | senior professional/technical worker (doctor, professor, lawyer, architect, engineer) |
| 02 | junior professional/technical worker (midwife, nurse, teacher, editor, photographer) |
| 03 | administrator/executive/manager (working proprietor, government official, section chief, department or bureau director, administrative cadre, village leader) |
| 04 | office staff (secretary, office helper) |
| 05 | farmer, fisherman, hunter |
| 06 | skilled worker (foreman, group leader, craftsman) |
| 07 | non-skilled worker (ordinary laborer, logger) |
| 08 | army officer, police officer |
| 09 | ordinary soldier, policeman |
| 10 | driver |
| 11 | service worker (housekeeper, cook, waiter, doorkeeper, hairdresser, counter salesperson, launderer, child care worker) |
| 12 | athlete, actor, musician |
| 13 | others |

Table A2

Appendix 2: Effect of Robot Exposure on Physical Health (Probit)

|  | (1) | (2) | (3) | (4) |
|---|---|---|---|---|
| *Dependent variable:* | *Illness or Injury* | *Muscle Pain* | *Rash* | *Cough* |
| Robot Exposure | -44.837*** | -395.810 | -146.606 | -45.026** |
|  | (15.168) | (404.695) | (157.766) | (20.191) |
| *Marginal Effect of Robot Exposure* | *-6.962*** | *-23.987* | *-4.050* | *-5.453*** |
|  | *(2.355)* | *(24.568)* | *(4.374)* | *(2.446)* |
| Observations | 3,296 | 2,665 | 1,076 | 2,678 |
| Control Variables | YES | YES | YES | YES |
| Year FE | YES | YES | YES | YES |
| Prefecture FE | YES | YES | YES | YES |
| Primary occuption FE | YES | YES | YES | YES |
| Employer ownership FE | YES | YES | YES | YES |

Robust standard errors in parentheses

*** p<0.01, ** p<0.05, * p<0.1

Table A3

Appendix Table 3: Effect of Robot Exposure on Physical and Mental Health （IV-2SLS)

|  | (1) | (2) | (3) | (4) | (5) |
|---|---|---|---|---|---|
| *Dependent variable:* | *Illness or Injury* | *Muscle Pain* | *Rash* | *Cough* | *Mental Health* |
| Robot Exposure | -15.355*** | 1.733 | -1.909* | -26.525*** | 11.384* |
|  | (4.215) | (2.165) | (1.152) | (6.592) | (6.242) |
| Observations | 3,361 | 3,370 | 2,805 | 2,808 | 2,371 |
| Demographic Controls | YES | YES | YES | YES | YES |
| Year FE | YES | YES | YES | YES | YES |
| Prefecture FE | YES | YES | YES | YES | YES |
| Primary occuption FE | YES | YES | YES | YES | No |
| Employer ownership FE | YES | YES | YES | YES | No |
| First stage (Robust F-value) | 144.78 | 145.241 | 279.051 | 218.644 | 1258.8 |
| Tests of endogeneity (p-value) | 0.0064 | 0.0057 | 0.6507 | 0.0002 | 0.0834 |

Robust standard errors in parentheses

*** p<0.01, ** p<0.05, * p<0.1